\documentstyle[12pt]{article}
\newtheorem{theorem}{Theorem}

\begin{document}
  \author{Uwe Brauer$^*$, Edward Malec$^{\#+}$ and
Niall \'O Murchadha$^{*+}$
  \\[0.52cm]
  $^*$Max-Planck-Institut f{\"u}r Astrophysik,\\
  Karl-Schwarzschild-Str. 1, \\
  D-8046 Garching b. M{\"u}nchen, FRG\\[0.2cm]
 $^{\#}$
Institute of Physics, Jagellonian University \\
  30-064 Krak\'ow, Reymonta 4, Poland\\ [0.2cm]
   $^{+}$ Physics Department, University College \\
   Cork, Ireland.}
 \title{Trapped surfaces in  spherical\\
 expanding open universes.}
\date{5/11/1993}
\maketitle

\centerline{PACS numbers: 04.20., 98.80., 95.30., 97.60.}

\begin{abstract}

Consider spherically symmetric initial data
for a cosmology which, in the large, approximates an open
$k = -1 ,\Lambda = 0$ Friedmann-Lema{\^\i}tre universe.
Further assume that the data is chosen so that the
trace of the extrinsic curvature is a constant and that
the matter field is at rest at this instant of time. One
expects that no trapped surfaces appear in the data if
no significant clump of excess matter is to be found.
This letter confirms this belief by displaying a
necessary condition for the existence of trapped
surfaces.This necessary condition, simply stated, says
that a relatively large amount of excess matter must be
concentrated in a small volume for trapped surfaces to
appear. \end{abstract}

 Over the last few years, we have been
investigating spherically symmetric solutions to the
Einstein equations, both asymptotically flat
\cite{BMN1} \cite{BMN2} \cite{BMN3}
and cosmological \cite{ICH3}, \cite{MNO},
with the intention of relating the
amount of matter inside some chosen sphere to the
possibility of trapped surfaces forming. The only
situation which has not yet been dealt with is the case
where we have a cosmological model with unbounded spatial
slices which, on average, has negative scalar
curvature. These are generalizations of the ``open''
(hyperboloidal) Friedmann universes. In this article we
will derive a necessary condition for the appearance of a
trapped surface in such a cosmology. This is very similar
to equivalent necessary conditions we have derived in all
the other cases. Interestingly enough, we have failed,
in this open case, to find the matching sufficient
condition which we were able to derive in all the other
cases. We comment further on this at the end of
this article.

Initial data for solutions to the Einstein
equations consist of four objects (see \cite{MTW})
[rack$ g_{ab}$, $K_{ab}$, $\rho$ (energy density) and
$J_b$ (matter current density)]rack;
where  $g_{ab}$ is a three dimensional, Riemannian
(positive definite) metric which describes a three
dimensional manifold $\Sigma$; $\Sigma$ is to be regarded
as a spacelike slice through the four manifold which is
the solution of the Einstein equations and $K_{ab}$ is a
symmetric three tensor which is the extrinsic curvature
of $\Sigma$ as an embedded surface in the four-manifold.
Consider a spacelike two surface $S$ embedded in $\Sigma$.
If the surface is orientable, we can identify a unit
vector $n^a$ as the ``outgoing'' direction orthogonal to
$S$ in $\Sigma$. Now $J_b n^b$ measures the outflux of
matter through $S$.

These data are not independent. They must satisfy the
constraints,
 \begin{eqnarray}
	^{(3)}R[g]  -  K_{ab} K^{ab}  +
   \left(K^{a}{}_a\right)^2   = 16\pi\rho
\label{1}
      && \;\;\mbox{(Hamiltonian constraint)}  \\
      D_a K^{a}{}_{b}
      - D_b K^{a}{}_{a} = -8\pi J_b             \label{2}
      &&\;\;\mbox{(momentum constraint)}
\end{eqnarray}
  where $^{(3)}R[g]$ is the scalar
curvature of $\Sigma$, $D$ the covariant derivative
compatible with $g_{ab}$. The trace ($g^{ab} K_{ab}$) of
the second fundamental form $K_{ab}$ is equal to the
(positive) rate of  change of the 3-dim volume in a
timelike direction $ \frac{d}{dt}(dV) =  g^{ab}K_{ab} dV$.

  Here we wish to consider initial data for
a spherically symmetric cosmology which in the large
looks like an open Friedmann universe. The standard time
slice through an open (hyperboloidal)
Friedmann-Lema{\^\i}tre universe is given by  a
hyperboloid of constant negative  curvature where the
line-element is given by
\begin{equation}
	d\widehat s^2 = a^2(t)  \left(	dr^2 + \sinh^2r
[		d\theta^2 + \sin^2\theta d\varphi^2  ] \right)
\label{3}    \end{equation}
where $a$ is constant on  $\Sigma$; $\widehat K_{ab}$ is
pure trace $\widehat K_{ab} = H \widehat g_{ab}$,
 $H=\frac{d
a}{dt}$/a is  the  Hubble constant; the (background)
energy-density $\widehat\rho$ is a constant, $\widehat
J^a = 0$. We will use this data as a background
against which we will compare our actual spherical
solution.   Background quantities will be denoted
  by a hat.

  The Hamiltonian constraint (\ref{1}) now gives
\begin{equation}
	-\frac{6}{a^2} + 6 H^2 =
	16\pi\widehat\rho \label{4} \end{equation}
while the momentum constraint (\ref{2}) is identically
satisfied.

 We wish to consider a spherically symmetric spacetime
whose spatial topology agrees with that of the open
Robertson-Walker geometry\footnote{If we refer to the
geometrical (or kinematical) properties of this spacetime
we use the term  Robertson-Walker, while for dynamical
aspects we use the term Friedmann-Lema{\^\i}tre.}. We
focus on a spacelike slice through the spacetime which
is itself spherically symmetric and on which the trace of
the extrinsic curvature is a constant. This is not as
restrictive a condition as it appears. If one is given a
spacetime with a spatial symmetry, the $trK = const.$
slices respect this symmetry \cite{NOM}.

Since the three-metric is spherically symmetric we know
that it is conformally flat and the line element can be
written as
 \begin{equation}
	ds^2 = \phi^4(r)	d\widehat s^2		\label{5} \end{equation}
where $\phi\geq 0$.
We can write the extrinsic curvature (assuming only
spherical symmetry) as  \begin{equation}
	K_{ab} = H g_{ab} + \left(n_an_b \:-\:
\frac{g_{ab}}{3}\right)K(r) \label{6} \end{equation}
where $n^a$ is the unit radial vector.
We further assume $J^a = J(r) n^a$.

Since $H$ is assumed constant, only $K(r)$ appears in the
momentum constraint. As in the
previous
 articles, the momentum constraint can
be written as an explicit integral
   to give
\begin{equation}
	K(R_S) =
	\frac{-12a\pi}{\phi^6(R_S)\sinh^3(R_S)}
	\int\limits_0^{R_S}
	\phi^8(r)\sinh^3(r) J(r) dr. \label{7} \end{equation}
 One immediate consequence of (\ref{7}) is that if the
matter is at rest in the slice ($J=0$) we have that $K=0$
and so the extrinsic curvature is pure trace. For the rest
of this article we will assume that this holds. Therefore
the momentum constraint is automatically satisfied and
we need not consider it further.

To evaluate the Hamiltonian constraint we first need
to calculate the scalar curvature of the metric. To
do so we use the background metric $\widehat g_{ab}$
and that $\widehat R = -\frac{6}{a^2}$. We can use the fact
that $g_{ab} = \phi^4 \widehat g_{ab}$ to write
\begin{equation}
	^{(3)}R[g] = -8\phi^{-5}\widehat\Delta\phi
		-\frac{6}{a^2}\phi^{-4}. \label{9}\end{equation}
 Hence the Hamiltonian constraint reads
\begin{equation}
	\phi^{-5}\widehat\Delta\phi = -   2\pi\delta\rho
-\frac{3}{4a^2}	\left(	1- \phi^4	\right)\phi^{-4}
\label{10} \end{equation}
where $\delta\rho$ denotes $\rho - \widehat\rho$. Let us
stress here that we make no assumption that $\delta\rho$
is either small or has a fixed sign.

 We wish to consider whether such a system will
gravitationally collapse to form a black hole. Evidence
would be the appearance of trapped surfaces in the
initial data. Consider a spherically symmetric two
surface $S$ in the initial data set.
The expansion of outgoing null rays from this
surface $S$ is given by
\begin{equation} \theta = D_an^a + \left(g^{ab}-
n^an^b\right)K_{ab}, \label{12}\end{equation}
 where $n^a$ is the unit (spacelike) normal to this
surface. If $\theta<0$ we say that the surface is (outer)
  trapped. One of the singularity theorems of general
relativity \cite{ROGER}, \cite{HAE}
tells us that the appearance of trapped
surfaces is an indication that gravitational collapse
is occuring. The expansion of a spherical surface in the
  open Friedmann universe is given by
\begin{equation}
	\theta = \widehat D_a \widehat n^a + 2H =
		\frac{2\coth r + 2a H}{a};
	\qquad \widehat n^a = (a^{-1},0,0). \label{13}\end{equation}
Therefore if the universe is expanding ($H>0$) no
trapped surfaces exist.

 In the spherically symmetric but non homogeneous model we
are considering here it is easy to show that the
 expansion can be written as
\begin{equation}
	\theta =  D_a  n^a + 2H =	\phi^{-3}
\left(	4\phi' + 2\phi\coth r	\right)
		+2 H;
	\qquad n^a = (\phi^{-1}a^{-1},0,0)
\label{14}\end{equation}
  where $\phi'=\frac{d\phi}{dr}$.

To relate this expression to the matter content of the
universe one integrates (\ref{10}) in the physical space
over a sphere $S$ of coordinate radius $R_S$. Using
spherical symmetry, integration by parts and Gauss law
the left hand side gives \begin{eqnarray}
	\int\limits_V
	\phi^{-5}\widehat\Delta\phi	dV
&=&		\int\limits_V
	\phi^{-5}\widehat\Delta\phi
	\phi^6 d\widehat{V}		\nonumber\\
&=&
	4\pi \phi\phi' \sinh^2 r|_R
-	4\pi \int\limits_0^{R_S}
	\phi'^2 \sinh^2 r a dr.	\label{15}
\end{eqnarray}
To do this integration one needs to use that $dV =
\phi^6 d\widehat{V} = \phi^6 a^3 \sinh^2r \sin\varphi$.
The right hand side of (\ref{10}) gives
\begin{equation}
   	\int\limits_V	\phi^{-5}\widehat\Delta\phi	dV =
   2\pi \delta M_S + \frac{3}{4 a^2}V
   - \frac{3}{4 a^2}\;
	4\pi \int\limits_0^{R_S}
	\phi^2 \sinh^2 r a dr.
\label{16}\end{equation}
  We equate (\ref{16})  with (\ref{15}),divide by $2\pi$,
recognize that the area $A$ of a sphere $S$ at a given
radius $R$ is given by $A=\int\limits_S \sqrt{|g|}d^2x
= 4\pi a^2 \phi^4 \sinh^2 R_S$ and use (\ref{14}) to
finally  get \begin{eqnarray}
	\frac{A}{8\pi}	\theta
&=&    -\delta M_S +	 \int\limits_0^{R_S}
	[	2\phi' \sinh^2 r  +			\left(	\phi^2 \cosh r \sinh r
		\right)'	-\frac{3}{2} \phi^2 \sinh^2 r
			] a dr	\nonumber\\ & &
		+\frac{3}{8\pi a^2}V	+ \frac{HA}{4\pi}
	\nonumber\\
&=& 	-\delta M_S +		 \int\limits_0^{R_S}
[		\frac{1}{2}		\phi^2  +		\frac{1}{2}
		\left(		\phi \cosh r
+		 2 \phi' \sinh r
		\right)^2 ]	a dr \nonumber\\ & &
		+\frac{3}{8\pi a^2}V	+ \frac{HA}{4\pi}. \label{17}
\end{eqnarray}
   (\ref{17}) is the appropriate equation to show a
   necessary condition;
   We assume $S$ to be trapped, i.e., $\theta<0$
   which leads to
\begin{eqnarray}
	0 &>&	 		-\delta M_S +		\int\limits_0^{R_S}
[		\frac{1}{2}		\phi^2
+		\frac{1}{2}		\left(	\phi \cosh r +		 2 \phi' \sinh r
		\right)^2 ]	a dr	+\frac{3}{8\pi a^2}V	+
\frac{HA}{4\pi}   \nonumber\\ &\ge&	-\delta M_S	+
\frac{1}{2}	L_S
		+\frac{3}{4 a^2}V	+ \frac{HA}{4\pi}.
\end{eqnarray}
The proper radius of the sphere of coordinate
 radius $R_S$ is given by
\begin{equation}
	L_S = \int\limits_0^{R_S} 	\sqrt{g_{rr}}dr
	= \int\limits_0^{R_S} 		\phi^2 a dr. \label{20}
\end{equation}
Hence, if a given surface is trapped, we get
\begin{equation}
		\delta M_S \geq	\frac{1}{2}	L_S
		+\frac{3}{8\pi a^2}V	+ \frac{HA}{4\pi}
\end{equation}

Finally we have:
\begin{theorem}[A necessary condition]\indent\\
Assume one is given initial data for a spherically
symmetric open cosmology, such that

{\setlength{\parindent}{1cm}
1.)  $K^{a}{}_{a} = const = 3H$;

2.)  $ J_b = 0$,}

and (background) Friedmann-Lema{\^\i }tre data ($a$,
$H$, $\widehat\rho$) satisfying eqn.(4).

If for a spherical surface
$S$, its proper radius $L_S$, its area $A$, the amount of
the excess mass $\delta M_S$ and the volume $V$ inside $S$
satisfy \begin{equation} \fbox{$\displaystyle
     \delta M_S < \frac{1}{2} L_S
     +   \frac{AH}{4\pi}
	+ \frac{3V}{8\pi a^2}$}
\end{equation}
then $S$ is {\bf not} trapped.
\end{theorem}

An obvious reorganization of Eqn.(19) can be achieved by
moving the term $\widehat\rho V$ from the left to the
right hand side. We then can use Eqn.(4) to give
\begin{equation}
\widehat\rho V + \frac{3}{8\pi a^2}V =
\frac{3H^2V}{8\pi}. \end{equation}
This gives us a condition for the nonappearance of
trapped surfaces which is independent of any choice of
background.
\begin{theorem}\indent\\
Assume one is given initial data as in Theorem 1.

If for a spherical surface
$S$, its proper radius $L_S$, its area $A$, the
total amount of matter $ M_S$ enclosed by $S$ and the
volume $V$ inside $S$ satisfy
\begin{equation}
\fbox{$\displaystyle
     M_S < \frac{1}{2} L_S
     +   \frac{AH}{4\pi}
	+ \frac{3H^2V}{8\pi}$}
\end{equation}
then $S$ is {\bf not} trapped.
\end{theorem}
This result forms the basis of a simple test which may
allow a determination of the global topology of our
universe. \cite{MON}

In the other cosmological models and in the case of
asymptotically flat data we have found essentially
identical necessary conditions and we have
also found sufficient conditions expressed
in terms of the same parameters.
Surprisingly, we have failed to date to
find, in the open case, a sufficient
condition. This should not lead one to jump
to the hasty conclusion that trapped
surfaces are absent from spherical expanding open
universes. It is easy to find
examples with trapped surfaces. Because of
the spherical symmetry, it is a simple matter to
construct a model which consists of a
constant density sphere embedded in an open
cosmology. If the density is large enough,
trapped surfaces appear.

\end{document}